# Would You Like to Motivate Software Testers? Ask Them How

Ronnie E. S. Santos
Centro de Informática – Universidade Federal de Pernambuco
Brazil
ress@cin.ufpe.br

Cleyton V. C. Magalhães
Centro de Informática – Universidade Federal de Pernambuco
Brazil
cvcm@cin.ufpe.br

Jorge S. Correia-Neto
Universidade Federal Rural de Pernambuco
Brazil
jorgecorreianeto@gmail.com

Fabio Q. B. da Silva
Centro de Informática – Universidade Federal de Pernambuco
Brazil
fabio@cin.ufpe.br

Luiz Fernando Capretz
Electrical & Computer Eng. Dept.
Western University
Canada
lcapretz@uwo.ca

Rodrigo E. C. Souza
Motorola Mobility Paternship (CIn/Morotorla)
Brazil
recs@cin.ufpe.br

*Abstract*—**Context.** Considering the importance of software testing to the development of high quality and reliable software systems, this paper aims to investigate how can work-related factors influence the motivation of software testers. **Method.** We applied a questionnaire that was developed using a previous theory of motivation and satisfaction of software engineers to conduct a survey-based study to explore and understand how professional software testers perceive and value work-related factors that could influence their motivation at work. **Results**. With a sample of 80 software testers we observed that software testers are strongly motivated by variety of work, creative tasks, recognition for their work, and activities that allow them to acquire new knowledge, but in general the social impact of this activity has low influence on their motivation. **Conclusion**. This study discusses the difference of opinions among software testers, regarding work-related factors that could impact their motivation, which can be relevant for managers and leaders in software engineering practice.

*Keywords—Software Testing, Motivation, Software Engineering*

I. INTRODUCTION

For decades, researchers have been interested in investigating practices to improve work performance of individuals in several different contexts [24]. During this period, many theories and discussions were developed aiming to enlarge the body of knowledge about this theme and contribute to the improvement of industrial practices. Researchers such as Viteles [31] and more recently Morgeson and Humphrey [22] emphasized the importance of understanding aspects of work design due to the increasing use of teams in the work place, while Hackman and Oldham [14] focused on the experience of workers and the individual work practices as one of the main elements to increase performance and the interest and attractiveness of the job.

Recently, human factors have become of great interest in the software engineering field as well, due to the fact that software development process commonly depends on human-centered activities [25]. In this context, several aspects of individual and team work in software development have been studied, seeking to understand particularities of human aspects in software engineering practice, such as the work design in software engineering [7], motivation of professionals [10][12][13], personality of individuals [1][6], work behaviors [9] and many others aspects that can directly impact the success of software development.

Some of these studies were consolidated in theories, for instance, there is a theory regarding the motivation and satisfaction of software engineers, hereafter referred as TMS-SE, that was developed based on the analysis of years of published field studies and from specific traits of software engineering practice [10][11]. This theory was proposed to support academic and industrial practice regarding the understating of motivation of software engineers in general, considering all professionals working in the software development process as a whole, that is, analysts, developers, testers, managers, and others. However, when analyzing the particularities of the different roles involved on the software development process, some studies raise discussions on the existence of peculiarities regarding professional, individual and work characteristics, and therefore, observable differences among those participating in this process[1][2][8].

Recently da Silva et al. [7] have discussed differences related to how professionals working with software development can experience different levels of interaction with work-related factors, such as motivation, satisfaction and burnout, depending on the role and the tasks performed in this process. These evidence demonstrated the importance of investigations about human factors and work characteristics not just in software engineering as a whole, but also, in each role and phase of the software development process.

Regarding this need for studies considering each specific phase of the software development process, previously, Kanij et al. [17] [18] have discussed the lack of evidence about human factors in software testing. These authors observed that the current research on this topic has focused mainly on the development of testing methodologies and tools and rarely discussed issues around human factors affecting professional software testers. This study pointed out the importance of visualizing software testing activities as a set of human dependent tasks and emphasized the need for research with critical interpretations about personal characteristics of software testers affecting the software testing process.

Considering the importance of software testing to the development of high quality and reliable software systems [23][26] and the lack of empirical evidence about human aspects affecting this activity [17], we decided to apply the definitions presented on the TMS-SE to answer the following question: *RQ. How can work-related factors influence the motivation of software testers?* We believe that this is an important topic to be explored in the industry practice because: a) motivation is an important factor that can directly affect the individual performance at work [14][15]; b) there is a theory (TMS-SE) [10] developed in the context of software engineering that can guide the research of motivation on the specific context of software testing.

To answer this research question we developed a cross-sectional survey-based instrument based on the TMS-SE and collected impressions and opinions from 80 software testers working at 3 different software companies. The TMS-SE defines motivation as the desire to develop a specific work, that is, the reasons that stimulate software engineers to perform their work, which can influence their productivity and the results of their work [10]. The theory argues that motivated software engineers in general are engaged and concentrated and states that there are five factors related to the work itself that can influence this motivation: *Acquisition of useful knowledge*, *Social impact*, *Work variety*, *Creativity* and *Well-defined work*. In this study we investigated how software testers perceived and valued these factors, presented in the theory as important to all individuals working in software development. Further, we explore the existence of new factors directly related to software testing. This effort can help industry practice to understand the impact of work-related factors on the motivation of these professionals, which can inform and support management and leadership in this context.

From this introduction, this paper is organized as follows. In Section 2, we present the conceptual background that characterizes this study. In Section 3, we describe the research method, instruments and techniques applied to answer our research question. In section 4 we present the main findings, which are discussed in Section 5. Finally, in Section 6, we present our conclusions and directions for future research.

## II. BACKGROUND

This section presents the theoretical background that supports this study, as well as related works in a similar context of this research.

### A. Motivation in Software Engineering

For a long time, the term *motivation* was used as a synonymous to *job satisfaction* and to several distinct behaviors of software engineers [10]. This disagreement among concepts represented a problem both to academic research and industrial practice, due to the need for the proper management of motivation in software companies, to achieve higher levels of productivity of professionals at work [13].

Nevertheless, as observed by Couger and Zawacki [5], and discussed by França, Sharp and da Silva [10] the existing theories developed in various contexts and commonly discussed in the literature, might not be completely applicable to software development environments, because individuals working with software development are part of a distinctive group of workers, considering their individual needs and, therefore, "what motivates software engineers is likely to be different from what motivates the population in general".

Regarding this particular problem, the TMS-SE brought light both for academic and industrial practice regarding the understating of motivation in software engineering, providing a theoretical framework with observable traits of motivated, not motivated, demotivated, and satisfied software engineers and discussed how important this understanding is to practitioners. Following this theory, there are five factors directly related to the work that could influence the motivation of software engineers [10] [11]:

- *Acquisition of useful knowledge*: the perception that the work provides knowledge gaining;
- *Social impact*: the perception that the work has impact on other lives;
- *Work variety*: the perception that the work is varied;
- *Creativity*: the perception that the work demands creative processes;
- *Well-defined work*: the perception that the work has a clear sequence of steps to be accomplished.

At this point, it is important to highlight that, following the theory, these factors can influence the motivation of software engineers in general, that is, all professionals working directly on the development of software. However, as discussed in [1][2][7][8] it is important to explore and investigate the particularities behind each phase of software development, and regarding human factors, the differences between each type of professionals involved in this process.

### B. Related Work

Regarding the main topic addressed in this research, we identified some studies about human factors in software testing activities. In this section we highlighted these studies and the findings related to the motivation in software testing. In fact, many researchers have discussed the importance of human aspects during this phase of the software development process [3][4][6][16]**Error! Reference source not found.**[20], and regarding the motivation of software testers the researchers were interested not only in understanding how these professionals feel about their work, but how and why they choose this specific career[32][33].

More than a decade ago, Weyuker et al. [33] observed that the most skilled software testers were accustomed to change jobs in the companies and become programmers, analysts, or system architects, because a career on software testing was not considered advantageous enough for most of the professionals. This scenario provides a wide variety of interpretations and questionings such as: "how demotivated has to be a software tester to abandon the career and follow other path in software development process?" To answer this question, it is important to consider that motivation is an antecedent of satisfaction [10], which has strong co-relation to job burnout, one of the main factors that can lead individuals to turnover [7].

Nowadays, the scenario seems to be similar. Recently, Waychal and Capretz [32] investigated the reasons why computer engineering graduates are not interested in testing careers. The findings indicate that students expect that their job will provide some elements that, following recent theories, are antecedent of work motivation, such as *Acquisition of knowledge*, *Creativity* and *Work variety*, and these individuals believe that a career in software testing might not provide this. Although those students that are interested or partially interested in a career in software testing visualize *Acquisition of knowledge* as a factor related to the work with testing tasks, the major percentage of individuals that participated of the study believe that this activity offers low levels of challenges and creative opportunities. Besides, part of the individuals see the testing process as tedious and repetitive, which compared to the statements of the TMS-SE could mean lack of *Work variety*, other important antecedent of motivation.

Although the study of Waychal and Capretz [32] is not centered in the problem of motivation of software testers, the findings indicate that most of the elements valued by students when choosing their career in software development are related to how motivated they expect to be in their work. In fact, a recent research of Deak et al. [8] investigated how professional software testers can be motivated at work, and pointed out the lack of challenges, variety of work, recognition and good management in software testing as central problems. These elements were considered essential towards the increase on the motivation of software testers.

So far, previous studies focused or related to the issue of motivation in software testing confirm that individuals working or expecting to work as software testers consider, as important, a set of elements strongly related to the type of work being performed. However, the evidence gathered so far do not demonstrate the level of importance that software testers attribute to each factor. The present study aims to contribute to the discussion about motivation in software testing, by adding new evidence to the body of knowledge of this specific topic, using an industrial survey-based approach to collect, analyze and synthesize opinions from these professionals.

III. METHOD

In this study we followed the guidelines of Kitchenhan and Pfleeger [19] and Linaker et al. [21] to perform an industrial cross-sectional survey. In this type of study, participants answer questions about one topic or phenomenon in one fixed point in time and the information can provide a snapshot of the context that is being studied.

*A. Desinging the Survey Instrument*

The questionnaire was designed to collect opinions from software testers about how the factors related to work affect their motivation. Following the guidelines, designing a questionnaire for a survey research require a team composed of experts with both research and domain expertise, which might provide both technical and practical knowledge about the topic under investigation [21]. In this sense, the questionnaire was constructed by two researchers, with previous experience as software testers in industry, and reviewed by an academic researcher (PhD professor).

Accordingly to the definitions of Kitchenhan and Pfleeger [19] to construct the survey questionnaire, we searched the relevant literature regarding the main research question, and built the instrument re-using part of previous instruments developed in the context of motivation in software engineering [13][17] and also based on the five assertions presented on the TMS-SE about the motivational factors related to the work itself [10][11]: *Work variety, Acquisition of knowledge, Creativity, Social impact* and *Well-defined work*.

To elicit the software testers' opinions, the survey included both closed and open questions. The instrument was designed in order to collect responses that could indicate how the definitions presented in the TMS-SE regarding the motivation of software engineers in general are perceived and valued by software testers, and also, that could reveal the existence of new work-related factors characterized in the specific context of software testing. In this research the instrument was developed regarding to:

1) Demographic information that could characterize the participants of this study;

2) Qualitative data that could spontaneously describe the work of software testers and their motivations to work;

3) Quantitative data that could reveal how software testers perceive and value the factors presented in the TMS-SE and that influence the motivation of software engineers in general.

As recommended in the guidelines, a pilot questionnaire was tested and validated in order to identify problems with the questionnaire and responses. Thus, following the guidelines [19] we submitted the questionnaire to a group of 6 specialists in the software testing context, composed by 1 researcher, 3 software test managers and 2 software team leaders. The group of specialists analyzed and evaluated the instrument in order to comment the questions, the phrasing of each question or possible misunderstandings in the questionnaire, and add items that they judged important to the survey. This group also provided information regarding how the participants could be grouped for data analysis, establishing the distribution of participants by years of experience as beginner (0-2 years), intermediary (2-6 years) and experienced (more than 6 years). Besides, these specialists pointed out the importance of analyzing the answers based on the experience of participants with test automation techniques.

After validation and adjustments based on the considerations received from these specialists during the pilot phase, the final questionnaire was organized in three major groups of questions, written in Portuguese and presented below translated to English.

TABLE I. SURVEY QUESTIONNAIRE

| Group | Questions |
|---|---|
| *Group I – Demographics: Questions that characterize the respondents and their experience with software testing* | Q1 – Gender |
| | Q2 – Highest level of education |
| | Q3 – Years of experience in software testing |
| | Q4 – Experience with software test automation |
| *Group II – Questions designed to assess factors related to work that influence the motivation of software testers* | Q5 (Open) – In general, how do you define your daily activities with software testing? Please, describe the tasks that you perform. |
| | Q6 (Open) – Please, characterize among the tasks that you perform those that you consider stimulating for you |
| *Group III – Questions designed to test the TMS-SE and explore factors presented on the theory that influence the motivation of software testers* | Q7 (Closed) - Given the following affirmations, select those that can characterize your daily activities as a software tester<br><br>( ) In general, I work with tasks that allow me to acquire new knowledge<br><br>( ) In general, I work with tasks that requires me to be creative<br><br>( ) In general, I work with tasks that have impact on other people´s life<br><br>( ) In general, I work with a set of different and variable tasks<br><br>( ) In general, I work with a set of well-defined and specific tasks |
| | Q8 (Closed) - Given the following affirmations, select the one that you consider the most important to you continue working with software testing or that you consider an important factor to your work<br><br>( ) Work with a diversity of tasks is the most important to me<br><br>( ) Work with tasks that allow me to learn new things is the most important to me<br><br>( ) Work with tasks with *social impact* is the most important to me<br><br>( ) Work with tasks that are well-defined from the beginning to the end is the most important to me<br><br>( ) Work with creative tasks is the most important to me |

*B. Procedure, Population and Sample*

We invited testers from two international software companies and one national company located in Brazil to participate in this study. About 185 software testers composed our population, considering that all individuals working in software testing activities in these three companies were invited to answer this survey, which included testers, developers-testers, QAs, test managers and leaders. In this study, we sent the survey to managers and team leaders of the companies by email and asked them to forward the invitation to their software testers.

The first company (Company A) is characterized as a test center that holds a partnership with an international mobile phones company. By the time of data collection the company had about 90 software testers working with several different tests approaches, such as regression, smoky and sanity, performance, usability, internationalization and location, and acceptance. The company is responsible for the implementation of the tests of products developed in USA and China, and the software testers have direct contact with the international branches of the main company.

The second company (Company B) also holds a partnership with an international mobile phones company. However Company B does not develop exclusively testing activities, since the partnership includes also the development of new products. By the time of data collection, the company had over 70 professionals, with 15 individuals working specifically with software testing.

The third company (Company C) is a private software organization specialized in software development and innovative software solutions in several business domains, such as finance, telecommunication, government, industry, services, and energy. By the time of data collection the company had over 500 employees working in 4 different cities in Brazil. We invited software testers working with the projects under development in the head office of this organization, which represents a group of about 70 individuals.

Considering the number of software testers working in the three companies, we can estimate a population of at least 185 professionals that received our invitation to participate of this study. By the end of data collection our sample was composed of 80 individuals, which represents a response rate of 43%, which we assumed a good rate since these professionals were not obliged to participate (were volunteers).

*C. Data Analysis*

We applied both qualitative and quantitative analysis in this study. To analyze the textual data collected from open questions, we applied the process that involves labelling and coding [29] the quotations provided by the respondents (Figure 1).

In this process, we compared the participants' responses to the definitions of motivational factors related to work presented in the theory [10], and also searched for factors not presented or not identified in the theory. In other words, the theory was applied in the qualitative analysis to provide definitions of factors related to the motivation of software engineers and previous discussed in the context of software engineering in general. Nevertheless, we also searched for factors that were not presented in this theory and that could be understood as a particularity in software testing.

On the other hand, the answers of closed questions were analyzed using descriptive statistics in order to present the distribution and frequency of individual values and believes exclusively related to the five antecedents of motivation described in the theory. In this phase, the data were explored with support of MS Excel™, which was also used to generate graphics and tables.

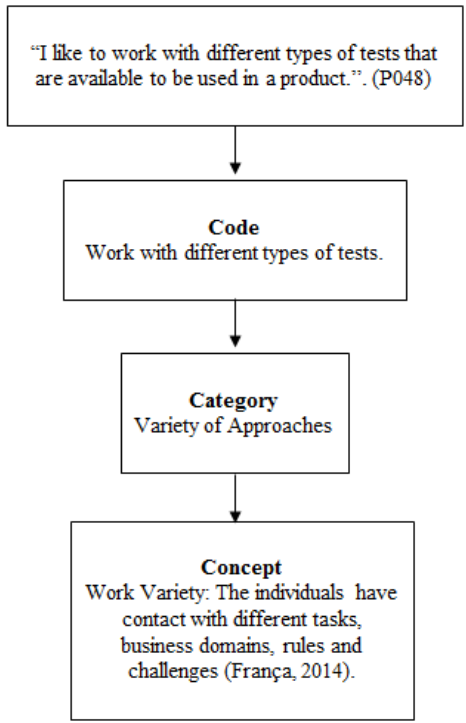

Figure 1. Qualitative Analysis Process

IV. RESULTS

We start this section presenting a brief description of the sample of individuals that participated of this study, and then we present the summary description of the answers to the survey questions.

A. *General Characterization of the Sample*

The survey received answers from 80 software testers, which represent 43% (80/185) of the studied population. In the sample, there was a prevalence of male individuals (67% or 54/80), something typical in this industry, while 33% (26/80) were female individuals. These individuals may be grouped based on their highest level of education, in which 65% (52/80) of the sample has a major degree in Computer Science, Computer Engineering or a similar graduation. Besides, almost 24% of participants (19/80) have a diploma or specialization in addition to their graduation. The remaining sample, 6% (5/80) have a master degree, 2% (2/80) have a PhD degree and 2% (2/80) have a technical degree.

Regarding the professional experience, about 36% (29/80) of the individuals in the sample were labelled as *beginners* in software testing, with less than 2 years of experience. Following this distribution, a similar percentage of the sample (28 individuals or 35%) had an intermediary level of experience, between 2 and 6 years working as software tester, and were labelled as *intermediary experience*. Finally, almost 29% (23/80) of the sample had more than 6 years of experience with software testing in industry, and were labelled as *experienced*.

Considering the testing approach commonly used in the work, 100% of the sample declared to have experience in performing manual testing. Considering automation, 80% of individuals (64/80) had previous or current experience working with automation in software testing, while 20% of professionals in the sample (16/80) have no experience with test automation.

B. *Open Questions: Exploring Motivation in software testing*

The second group of questions presented in the survey was designed to assess what work factors are part of the work of software testers and have impact in their motivation. Through this group of questions the individuals could spontaneously comment about their work and describe their everyday tasks without considering or knowing any previous concept presented in the theory. We analyzed this set of answers applying the method described in Figure 1 and identified 5 factors directly related to work that can influence the motivation of these professionals, 4 factors presented in the theory and one new factor, as summarized in TABLE II.

TABLE II. PERCEPTION OF ANTECEDENTS OF MOTIVATION (SPONTANEOUSLY)

| Antecedent | Perception Percentage |
|---|---|
| Work variety | 30% of participants |
| Creativity | 24% of participants |
| Recognition for work | 23% of participants |
| Acquisition of useful knowledge | 21% of participants |
| Well-defined work | 10% of participants |

From our sample, about 30% (24/80) of participants pointed out that working with software testing is commonly characterized by the diversity of tasks and the use of different approaches and technologies to perform the work. In this context, *Work variety* is an important factor related to the work of software testers and that can influence their motivation, as illustrated in the quotations below (translated to English by the researchers).

*"Every day I have to handle different issues, so it is not a boring work." (P033)*
*"I contribute to the construction of quality software, so I have different challenges every day." (P045)*

Further, 24% of participants (19/80) believe that software testing tasks demand *Creativity* to identify bugs and explore potential software issues and problems.

*"It's an activity more creative than it looks. It's not just following the steps." (P016)*

*"It is a creative, fun and enjoyable work."* (P054)

Following this process, software testers also considered the recognition for their work as an important motivational factor. We observed that 23% (18/80) of participants described software testing as a worthy activity, of great importance to the success of the software project and with visible accomplishments. Thus, the *Recognition of work* is another important factor that can influence their motivation. This factor is not described in the TMS-SE as a work-related factor that antecedes the motivation of software engineers in general, which means that this factor could be more related to software testing than other phases of the software development process.

*"The fact that you know that you are the responsible for the guarantee of quality in a product is stimulating."* (P005)

*"You see that your role aggregates value to the team, and that you are contributing to the software quality."* (P013)

Still regarding the open questions, 21% of participants characterized the work of software testers as a great opportunity to learn and acquire professional knowledge for the present and future opportunities of work (*Acquisition of useful knowledge*), which has an important role in their motivation.

*"You have the possibility to acquire knowledge about new technologies, like mobile, desktop or web."* (P025)

*"I always learn something new, useful to perform my work the best as I can."* (P014)

Finally, 10% of participants described software testing tasks as a *Well-defined work*, that is, a set of systematic tasks with clear goals and predictable results.

*"It's like a cycle, you find a problem, you report it and then you see that this issue was fixed. This is really good."* (P010)

Our analyses also demonstrated that individuals with more years of experience are those who see *Work variety* as the main work factor present in software testing activities, while those in the beginning of the career and with less experience believe that software testing is a process more related to *Creativity*. Participants in the beginning of their career also see better opportunities to learn as software tester than those with a higher level of experience. *Well-defined work* was a factor perceived in the same level by all the participants, considering this spontaneous approach to describe their work. We couldn't identify significant difference between the opinion of those who had and those who didn't have experience with software test automation.

Finally, it is important to report that 6% of the answers (5/80) were inconclusive or did not provide sufficient data to be analyzed. For instance, participants that answered the question with "I don't know" or only citing tools that were used in their work.

By this point, we have gathered information about work-related antecedents of motivation specifically related to the software testing process, since the professionals had spontaneously commented and described their work and the activities that stimulate them to work. Thus, the next step was to understand how their perception can be expressed considering the definitions presented in the TMS-SE. Thus, in the third group of questions, we created a set of alternatives, each one describing one of the five antecedents of motivation related to the work and described in the TMS-SE, and asked participants to select those that could characterize their work and those that are more responsible for stimulate them at work.

*C. Testing the TMS-SE in software testing*

When participants were asked about the general characterization of their work as software testers, by selecting options constructed based on definitions of TMS-SE, all five antecedents of motivation directly related to the work were considered as existent in software testing activities. However, following the participants' perceptions, not all these factors are strongly representative in this specific work, as presented in TABLE III.

TABLE III. PERCEPTION OF ANTECEDENTS OF MOTIVATION (TMS-SE BASED)

| Antecedent | Perception (Spontaneous) | Perception (Induced) |
|---|---|---|
| *Work variety* | 30% of participants | 66% of participants |
| *Creativity* | 24% of participants | 76% of participants |
| *Acquisition of useful knowledge* | 21% of participants | 84% of participants |
| *Well-defined work* | 10% of participants | 63% of participants |
| *Social impact* | 0% of participants | 44% of participants |

By answering questions based on the definition of concepts in the literature, we observed an increase in the level of perception of all factors. In this scenario, 87% of software testers (67/80) indicated that working with software testing provides them opportunity to acquire new knowledge (*Acquisition of useful knowledge*). *Creativity* remains as the second factor more perceived by this group of professionals, with 76% of answers (61/80), followed by *Work variety* with 66% of answers (53/80). *Well-defined work* had 63% on answers (50/80) and *Social impact* appears as the less representative, being perceived in the work for about 44% of participants (35/80).

When considering the experience of participants (years of experience) and the answers to this question, we couldn't identify any significant difference to *Acquisition of useful knowledge*, *Creativity* or *Work variety*. However, considering *Well-defined work* we observed that participants at the beginning of the career tend to see software testing as a set of systematic and well defined tasks, more than those that are working in the field for a longer period. We also observed that over half of the individuals at the beginning of their career as software testers see software testing as an activity with considerable *Social impact* over other people's lives, while only 30% of individuals with more than 6 years of experience have the same perception.

Although all the antecedents of motivation directly related to the work were perceived by the participants as present in the software testing, the difference between spontaneous and induced questions show that a great number of participants do not see *Social impact* as a factor strongly related to their work. Thus, at this point, we applied the definitions presented in the TMS-SE to assess the level of importance of each of the five factors. That is, how software testers value the antecedents of motivation that in theory, are related to the general work in software development. Therefore, participants were invited to indicate how they see the importance of these antecedents of motivation related to the general work in software engineering.

As summarized in TABLE IV. *Acquisition of useful knowledge* is the factor related to the work that can have more influence on the motivation of software testers, since almost half of the participants (48% - 38/80) had considered it as the most important factor related to their work. Following this, 20% of software testers (16/80) considered more important to know exactly what are the goals, the requirements and the results of the tasks that they are being assigned to perform (*Well-defined work*). The third more important factor to these professionals is the *Work variety*, the possibility to work with different types of tests, domains and technologies, representing 14% of the answers (16/80). *Creativity* is the fourth factor in level of importance to software testers with 11% of answers (9/80). Finally, *Social impact* can be considered as the less important factor in software testing with only 8% of answers (6/80).

TABLE IV. LEVEL OF IMPORTANCE OF ANTECEDENTS OF MOTIVATION RELATED TO WORK

| Antecedent | Perception |
|---|---|
| Work variety | 16% of participants |
| Creativity | 11% of participants |
| Acquisition of useful knowledge | 48% of participants |
| Well-defined work | 20% of participants |
| Social impact | 8% of participants |

When analyzing the data considering the experience of participants (years in industry) we observed few differences in how these professionals value the factors. *Acquisition of useful knowledge* is the most important factor to all three groups of professionals (Beginners, Intermediary Experience and Experienced). Nevertheless, we observed that experienced professionals value more *Work variety* than *Well-defined work*, different from those professionals in the beginning of the career or with an intermediary level of experience. Few variances can also be observed considering the level of importance of *Creativity* and *Social impact*. Figure 2 illustrate the difference in the answers of professionals regarding their period of experience in software industry.

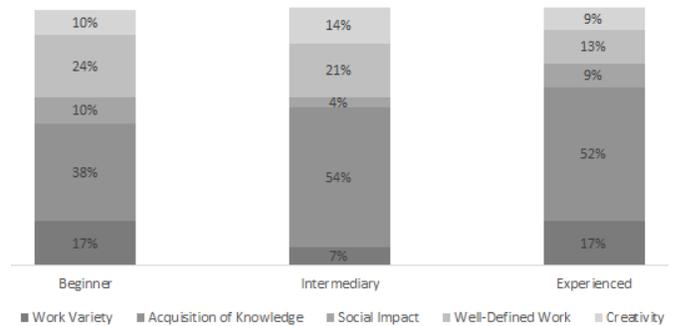

**Figure 2. Importance of Factors by years of experience**

Further, considering the experience with test automation, we observed that the general scenario is similar to participants that have worked or are working with techniques of automation, regarding the level of importance of each of the five factors. However, to those participants with no experience with automation, there is a considerable change in the value of these factors. This group of participants does not consider *Work variety* as an important factor, and they consider *Social impact* more important for their motivation than *Creativity*. This information is summarized in Figure 3.

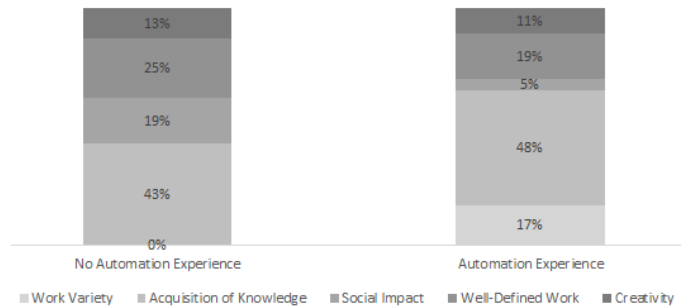

**Figure 3. Importance of Factors for professionals with experience with test automation**

## V. DISCUSSIONS

In this section, we discuss the findings of our research. We start comparing our findings with the outcomes of previous studies. Then we discuss the implications of these results for research and practice.

### A. Comparing Evidence with Previous Studies

As discussed in Section 2.1 the theory of motivation and satisfaction of software engineers was constructed based on data collected from four case studies with the participation of different types of professionals working in software companies. In general the theory states that, regarding the general work in software engineering, there are five factors that could influence the motivation of professionals: *Work variety*, *Acquisition of useful knowledge*, *Creativity*, *Well-defined work* and *Social impact*. However, our results shows that software testers, in particular, usually do not perceive the existence of these five factors.

In fact, when these professionals are asked to describe and comment about their work, there is one factor that is ignored, the *Social impact*. Nevertheless, when the same group is asked to select the factors from a previous list, this element is considered as part of their job, but, for a small percentage of professionals. Based on our findings, we can argue that regarding the motivation of software testers, managers and team leaders should provide regular opportunities to acquire new knowledge, and should define strategies to increase the levels of *Work variety*, since these are the two most important work-related factors in this context. Besides, these practitioners should develop techniques to manage the levels of the remaining of factors related to work that could impact the motivation of software testers, namely, *Creativity* and *Well-defined Work*.

Our results demonstrated that software testers consider the recognition of their work as an important factor that can influence motivation. *Recognition of work* is a new factor identified in this study. Based on our analysis, we encourage managers and leaders to develop strategies to let software testers know that their effort is appreciated. This attitude can have a lasting impact on their motivation. Besides, discussions in the literature demonstrate that the work recognition has an impact beyond the motivation, such as, increase of individual productivity, teamwork enhancement, and retention of quality professionals, reduction of turnover through job satisfaction increase, and reduction of absenteeism and stress, which can avoid job burnout.

Further, we can argue that the level of experience of software testers can moderate the value that these individuals attribute to the general antecedents of motivation related to the work. The results point out that the importance that software testers attribute to the antecedents of motivation may vary over the years, and professionals in the beginning of career may overvalue some factors, such as *Well-defined work*, while professionals with more experience seem to overlook the same factors and have their motivation more affected by the possibility of acquiring new knowledge, for example. Therefore, we can confirm that the balance of these elements can be efficient to obtain good results on motivation of software testers.

When analyzing the experience of software testers with test automation, we observed that, contrary to the majority, professionals with no experience with test automation approaches tend to ignore the need of *Work variety* in their job, which could imply a problem since this factor is an important component of *Engagement* [10]. We still need more information to discuss and understand this specific scenario. Nevertheless, we have reasons to believe that this characteristic is related to the interest of these individuals for *Specialization* at work [28] and the work with consistent and well-defined steps. Previous studies about job rotation in software companies [27] [28] demonstrated that some professionals working with software development have a positive attitude towards becoming a specialist; therefore, they can reject the idea of being rotated among different types of projects or work, which would mean higher levels of *Work variety*. Further investigations can confirm or refuse this proposition.

*B. Implications*

This study presents a set of implications for software engineering research and practice. Our research shows that to be applied in industry, theories ought to be re-analyzed regarding particularities due to the different characteristics of roles and types of professional working in the software development process. In this specific case, the general definitions about the motivation of software engineers are here demonstrated in the specific context of software testing.

We investigated the use of a general theory constructed in the context of software engineering to understand how its statements can vary when considering a very specific group of professionals. Our results show that some of the general definitions about work-related antecedents of motivation discussed in the TMS-SE are valid in the context of software testing, such as *Work variety, Acquisition of useful knowledge*, *Well-defined work* and *Creativity*. However, we identified two main differences while investigating a specific context in comparison to the general context of software engineering. First, software testers might not be strongly affected by activities with lack of *Social impact*, as much as other professionals working in software development. Thus, we can hypothesize that a similar situation regarding this or other factors can be observed in different other phases of the software process, such as requirements or design, for example. Second, new factors might be identified in other studies, regarding different types of professional working with software development. In this study, the *Recognition of work*, not discussed in the TMS-SE as an element related to the work, is an important factor regarding motivation in software testing.

Further, regarding the recent findings of Waychal and Capretz [32] about the reasons for why computer engineering graduates are not interested in careers in Testing, our results demonstrate the importance of discussions with students, regarding the characteristics of different roles and professionals necessary to develop efficient and quality software. As observed by these authors, students could be less interested in software testing because they believe that the job cannot provide them with a variety of tasks, creative work and challenges. Our study, on the other hand, demonstrates that these elements are part of the work in software testing, and pointed out that these are important factors to the motivation of professionals working in this area.

Our results can help managers and team leaders that are dealing with motivational problems in software testing, since our discussions demonstrated which work-related factors are more important for software testers and which of them can strongly affect their motivation. Therefore, these discussions can guide industrial practice to handle demotivation issues that could affect software quality. As discussed before, software testing is a human-dependent activity in software development and the motivation of software testers can directly influence the quality of the final product.

*C. Threads to Validity*

Considering validity, we looked for consistency of our instrument by using a pilot study with specialists both from academy and practice. Thus, we believe that the data collected in this study demonstrated good consistency.

Regarding our sample, we understand that for future generalizations we need to increase the number of participants seeking for more variation of gender, experience with testing approaches and also regarding the type of software and organizational contexts. However, the evidence collected and synthesized so far is a starting point for discussions regarding the motivation in software testing.

One important threat to external validity in our sample is the fact that all participants are located in Brazilian organizations. However, we can argue that at least two of the three companies participating in the study have a partnership with large international companies, which means that the participants are accustomed to work with procedures and domains that are applied in other countries. Further, we intend to replicate the study with companies from other countries in order to generalize our findings in the future.

## VI. CONCLUSIONS

We developed a survey-based instrument to understand issues related to the motivation of software testers. This instrument was constructed based on a previous theory about motivation constructed in the context of software engineering. We collected opinions from a sample of 80 software testers about the characterization of the work in software testing and about the factors related to the work that influence the motivation of these professionals.

We observed that the statements of the theory about the antecedents of motivation related to the work in software engineering are partially applicable when considering the specific context of software testing, with few modifications and the addition of a new element. However, it is important to notice that the factors that motivate software engineers in general, can affect software testers in different levels. For instance, *Social impact* is a factor with low importance to software testers, while *Acquisition of knowledge* is the factor that has more impact on their motivation. Software testers, in general, are also motivated by *Work variety* and *Creativity*. Nevertheless, the level of experience of these professionals, both in years and also in techniques of automation, can moderate the effect of these factors.

We expect that these results may be useful for the industrial practice, in order to support managers and leaders while facing problems with motivation of software testers at work. Besides, this study briefly discusses the importance of discussions about the characteristics of work in software testing to students in order to direct good professionals to this phase of software development process.

In future studies we expect to investigate questions related to the difference on the motivation of professionals working with test automation and those with few or no experience with this approach. Besides, we intend to replicate this study collecting data from software testers in companies in different contexts (both industrial and cultural) in order to generalize our findings.

ACKNOWLEDGMENT

Fabio Q. B. da Silva holds a research grant from CNPq #314523/2009-0. Cleyton V. C. Magalhães and Ronnie E. S. Santos are PhD students and receive a scholarship from CNPq. We are also very grateful to all participants for dedicating their time and attention to our research.